# Stoichiometric growth of high Curie temperature heavily-alloyed GaMnAs

S. Mack, R. C. Myers, J. T. Heron, A. C. Gossard, D. D. Awschalom[a)]

*Center for Spintronics and Quantum Computation, University of California, Santa Barbara, California 93106*

**Heavily-alloyed, 100 nm $Ga_{1-x}Mn_xAs$ ($x$>0.1) films are obtained via low-temperature molecular beam epitaxy utilizing a combinatorial technique which allows systematic control of excess arsenic during growth. Reproducible, optimized electronic, magnetic and structural properties are found in a narrow range of stoichiometric growth conditions. The Curie temperature of stoichiometric material is 150-165 K and independent of $x$ within a large window of growth conditions while substitutional Mn content increases linearly, contradicting the prediction of the Zener Model of hole-mediated ferromagnetism.**

[a)] Electronic mail: awsch@physics.ucsb.edu

Greater understanding of defect formation is needed to further increase the Curie temperature ($T_C$) of ferromagnetic semiconductors, such as $Ga_{1-x}Mn_xAs$, which hold promise for magneto-electronic devices.[1] In typical molecular beam epitaxy (MBE) growth of GaMnAs, low substrate temperatures ($T_{sub}$) of ≈250 °C are required to suppress precipitation of secondary crystal phases (e.g. MnAs) and incorporate a large fraction of Mn atoms as substitutional ($Mn_{Ga}$) acceptors in GaAs. Mn atoms also form double-donor interstitials ($Mn_i$) that compensate holes. Annealing films in air at temperatures of ≈180 °C removes $Mn_i$,[2] thus increasing the hole density ($p$) and $T_C$, in agreement with the Zener model of carrier-mediated ferromagnetism.[3,4] This model also predicts that $T_C$ increases linearly with $x$, which is supported by studies over the last decade with $0.02<x<0.08$.[4] More recently, MBE growth of GaMnAs with $x\geq0.1$ was demonstrated in attempts to increase $T_C$..[5,6] These films require reduced $T_{sub}$ (≈180 °C) and are limited to ≤10 nm thickness; $T_C$'s after annealing reach the previous limit of ≈170 K.[4] In these studies, the effect of arsenic flux was not explored, though it should be a critical parameter at these conditions.

GaAs films grown at $T_{sub}$ ≈250 °C can contain double-donor arsenic antisites at concentrations $As_{Ga}$≈$10^{20}$ $cm^{-3}$ (≈1% of Ga sites).[7,8] For a fixed $As:Ga$ flux ratio, $As_{Ga}$ in low-temperature (LT) GaAs increases exponentially as $T_{sub}$ is decreased below ≈300 °C.[9,10] Extended defects are prominent below 200 °C leading to polycrystalline films.[11,12] Previously, we developed combinatorial (non-rotated) growth to continuously vary $As:Ga$ and, therefore, $As_{Ga}$, in GaMnAs films near the paramagnetic/ferromagnetic transition (≈1% Mn) with $T_{sub}$ =250 °C.[13,14] There, excess arsenic flux of ≈1% suppresses ferromagnetism, an unfeasible flux precision for rotated growths. Here we extend this technique to $T_{sub}$ ≈150 °C, where the metastable solubility limit of Mn in GaAs is increased beyond 10%. As in LT GaAs, increased $As_{Ga}$ and extended defects occur unless *As:Ga* is precisely balanced to achieve stoichiometry. We discuss systematic variations of magneto-transport, ferromagnetism and crystal structure of GaMnAs ($x \geq 0.1$) as *As:Ga* is varied along the [110] direction (Fig. 1a inset). Optimal material is obtained in a narrow band of stoichiometric material. Stoichiometric, 100-nm thick films exhibit excellent crystalline quality and square magnetic hysteresis up to 165 K. Stoichiometric films grown with varying $x$ (>0.1), $T_{sub}$ and growth rate reach the same $T_C$ (150-170 K) after annealing; increasing $Mn_{Ga}$ does not increase $T_C$ as the Zener model predicts. We find the linear dependence of $T_C$ on $Mn_{Ga}$ is limited to $0.02<x<0.1$.

Samples are grown on 2", semi-insulating (001) GaAs wafers by MBE as described in Ref. 13. The total fraction of Mn in the films ($x$) is calibrated from growth rate calibrations of MnAs and GaAs reflection high-energy electron diffraction (RHEED) intensity oscillations. After growth of the buffer layer, $T_{sub}$ is dropped to 150 °C and substrate rotation is stopped. Below 350 °C, the arsenic shutter is closed; it is opened for 10 s prior to LT growth to ensure an arsenic-terminated surface. $T_{sub}$ is measured in real-time during growth using band-edge thermometry with an accuracy of ±2 °C.[15] During the first 5-15 nm of growth, $T_{sub}$ increases by ≈10 °C due to radiation from the heated Ga and Mn sources; $T_{sub}$ quickly stabilizes after heater power is dropped. We observe a 2D (streaky) 1×2 RHEED pattern on As-rich material transitioning to a 3D (spotty) pattern on Ga-rich material, observed by eye as mirror-finish (As-rich) transitioning to haze from Ga droplets on the surface.[13] Wafers are cleaved into ≈3x3 $mm^2$ pieces along [110], yielding ≈17 samples with systematically varying *As:Ga*.

Magnetometry and magneto-transport are measured on sample pairs with equal *As:Ga* (adjacent in the [$\bar{1}$10] direction). Magneto-transport is measured in the van der Pauw geometry from 2-380 K with out-of-plane (hard axis) fields up to 14 T. $T_C$ is determined from superconducting quantum interference device magnetometry (Quantum Design MPMS SQUID VSM) while warming in 50 Oe ([110] in-plane) after cooling from 350 K in 1 T. Saturation moment ($M_{sat}$) is determined from hysteresis loops (<0.5 T) . We anneal sample pairs in air at ≈180 °C until conductivity is maximized (Fig. 1c inset).[2] High-resolution x-ray diffraction (HRXRD) is performed on half-wafers using a triple-axis (Phillips X'Pert MRD PRO) diffractometer with a 0.25 ° aperture for ≈1 mm resolution along [110]. Lattice constant ($a_{GaMnAs}$) is determined from (004) ω-2θ scans.

Electronic, magnetic, and structural properties all depend critically on *As:Ga*. Figure 1 plots data from a single 100-nm thick $Ga_{0.84}Mn_{0.16}As$ film grown with $T_{sub}$=150 °C at 0.852 Å/s. Room temperature longitudinal ($\sigma_{xx}$) and Hall ($\sigma_{xy}$) conductivity (Fig. 1a), $T_C$ and $M_{sat}$ at 5 K (Fig. 1b), and $a_{GaMnAs}$ (Fig. 1c) all maximize at the same *As:Ga* where stoichiometry is achieved. $\sigma_{xx}$ increases with annealing (Fig. 1c inset) due to out-diffusion of $Mn_i$ from the bulk to the surface.[2] Examples of ω-2θ scans in Fig. 1d reveal high-quality XRD epi-layer peaks with thickness fringes observed at stoichiometry (black points) which match dynamical simulations (line).[16] The post-annealing epi-layer shift (red) shows a reduction of $a_{GaMnAs}$ from removal of $Mn_i$. This corroborates the model of Masek *et al.*[17] and results of Sadowski and Domagala[18] demonstrating that the main contributor to $a_{GaMnAs}$ expansion is $Mn_i$. It follows that the largest number of $Mn_i$ occur for stoichiometric material, the same *As:Ga* at which $T_C$, $M_{sat}$, $\sigma_{xx}$, and $\sigma_{xy}$ are largest.

As-rich material displays a broad XRD peak without thickness fringes indicating low crystal quality (blue, Fig. 1d). At this *As:Ga*, ferromagnetic MnAs particles (NiAs structure) are detected by magnetometry ($T_C$ = 320 K). No remnant magnetization ($M_{rem}$) is observed at temperatures above the GaMnAs $T_C$ in the stoichiometric region, but a $M_{rem}$ of 0.1 $\mu_B$/Mn from MnAs persists above the GaMnAs $T_C$ in the film with 35% excess arsenic. Excess arsenic might promote MnAs precipitation by creating extended defects that act as nucleation sites. MnAs is not observed in thinner As-rich films (7.5 nm), which could be below the critical thickness for defect formation. For our As-rich material, the same limits to $T_C$, $\sigma_{xx}$, and film thickness as in Refs. 5,6 are reproduced; these limits are overcome by achieving stoichiometry at precise *As:Ga*.

Excess arsenic and $Mn_i$ significantly alter magnetic and electronic properties such as $T_C$, $M_{sat}$, coercive field ($H_C$), resistivity ($\rho_{xx}$) and magnetoresistance (*MR*). Figure 2a shows the $T_C$ increase to ≈165 K after minimizing defects, and we observe square hysteresis at this high $T_C$ (Fig 2b). Excess arsenic reduces the $H_C$ and $M_{sat}$ at 5 K (Fig. 2c). $\rho_{xx}$ and *MR* increase by orders of magnitude at 10 K in samples with $Mn_i$ and excess arsenic (Fig 2d). $Mn_i$ increases *MR* (at 14 T) from 3% to 44%, and excess arsenic (*As:Ga* =12) increases it to 87%. $\rho_{xy}$ switching occurs in all samples due to the anomalous Hall effect (Fig 2d).

The growth parameter space of heavily-alloyed GaMnAs is systematically explored with 16 non-rotated wafer growths with $0.1<x<0.22$, $120<T_{sub}<160$ °C, and growth rate varied from 0.198-1.41 Å/s. Figure 3a plots $T_C$ against $x$ for the stoichiometric samples from each heavily-alloyed wafer and also lightly-alloyed wafers ($x<0.1$, $200<T_{sub}<250$ °C). We observe the expected linear increase in $T_C$ with $x$ ($x<0.1$),[3,4] but it does not extend to the heavily-alloyed regime, where $T_C$ = 150-165 K post-annealing for all $x$ (maximum observed at $x$=0.16) despite the wide range of growth parameters. In contrast, $a_{GaMnAs}$ increases linearly up to $x\approx0.16$ for both as-grown and annealed films (Fig. 3b) and is fit using the model from Refs. 17,18. Assuming $As_{Ga}\approx0$ in the stoichiometric region, $a_{GaMnAs} = a_{GaAs} + 0.02(x\text{-}z) + 1.05z$, where $z$ is the atomic fraction of $Mn_i$ and ($x$-$z$) is the atomic fraction of $Mn_{Ga}$. Linear fits, plotted as lines in Fig. 3b, demonstrate a constant fraction of interstitials ($z/x$) of 0.24 in as-grown samples and a reduction of this fraction to 0.13 after annealing. The linear behavior indicates that $Mn_{Ga}$ and $Mn_i$ increase proportionally with $x$, up to $x \approx 0.16$. Contrary to the predictions of the Zener model, $T_C$ is independent of $x$ for $x>0.1$ even though $Mn_i$ compensation maintains the same proportions in this heavily-alloyed regime.

In conclusion, heavily-alloyed, 100–nm thick GaMnAs films ($0.1<x<0.22$) with reproducible, high magnetic ($T_C \approx 165$ K) and structural quality are grown utilizing a combinatorial technique to achieve stoichiometry. Magneto-transport, ferromagnetism, and lattice constant are critically dependent on $As:Ga$ for low $T_{sub}$, with stoichiometric, annealed material displaying optimal properties. While structural and magnetic data indicate a linear increase in $Mn_{Ga}$ up to $x \approx 0.16$, we do not observe the $T_C$ increase predicted in Ref. 3, suggesting the Zener model may not be applicable to the heavily-alloyed regime. Application of this combinatorial technique provides a reproducible method for obtaining high-$T_C$ GaMnAs and allows systematic exploration of the growth parameter space.

We thank J. H. English and A. W. Jackson for MBE advice and M. A. Scarpulla for helpful discussions. This work was supported by ONR, MURI and AFOSR. The authors used MRL Central Facilities supported by the MRSEC Program of the NSF Contract No. (DMR05-20415). S. M. acknowledges support by the DoD through the NDSEG Fellowship Program.

## Figure Captions

**Fig. 1.** Electrical, magnetic and structural dependence on *As:Ga* for a 100-nm thick $Ga_{0.84}Mn_{0.16}As$ film grown without rotation. **Inset:** substrate and source geometry. **(a)** Room temperature longitudinal ($\sigma_{xx}$) and Hall ($\sigma_{xy}$) conductivity, **(b)** Curie temperature ($T_C$), saturation moment ($M_{sat}$) at 5 K and **(c)** lattice constant ($a_{GaMnAs}$) for different *As:Ga*. Stoichiometric region is shaded grey. **Inset:** $\sigma_{xx}$ versus anneal time, right axis is anneal temperature. **(d)** HRXRD scans along ω-2θ near the (004) substrate peak measured on stoichiometric (as-grown and annealed) and As-rich samples labeled in the figure.

**Fig. 2.** Variation in ferromagnetism and magneto-transport due to stoichiometry and post-growth annealing (same film as Fig. 1). **(a)** Magnetic moment (*M*) and $\sigma_{xx}$ versus temperature (*T*). **(b)** *M* versus magnetic field (*H*) hysteresis loops from the stoichiometric, annealed sample at various temperatures near $T_C$. **(c)** Hysteresis loops at 5 K. **(d)** Resistivity ($\rho_{xx}$) at 10K versus *H*. Magneto-resistance (*MR*) at 14 T is labeled in the figure. Hall resistivity ($\rho_{xy}$) versus *H* for the annealed, stoichiometric sample.

**Fig. 3.** **(a)** $T_C$ versus *x* for stoichiometric samples of $Ga_{1-x}Mn_xAs$ for $0<x<0.22$ from many separate non-rotated growth runs. Data are shown both for as-grown and optimally annealed samples. Lines are guides to the eye. **(b)** Lattice constant ($a_{GaMnAs}$) versus *x* from HRXRD scans as shown in Fig 1d. Lines are fits as described in the text indicating a constant fraction of interstitials (z/*x*) labeled in the figure. Blue data are for increasingly As-rich material, along the arrow.

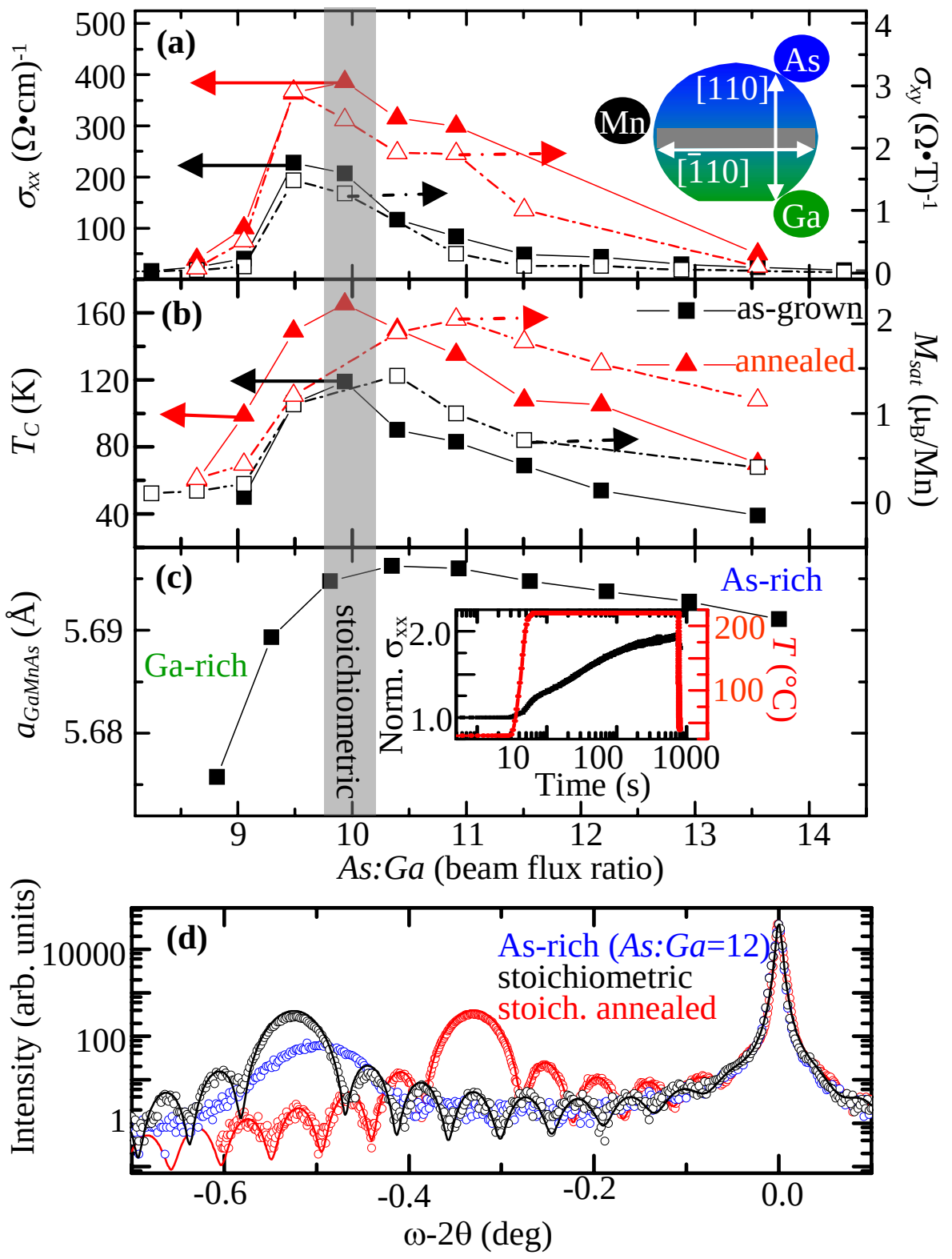

Figure 1
Mack et al.

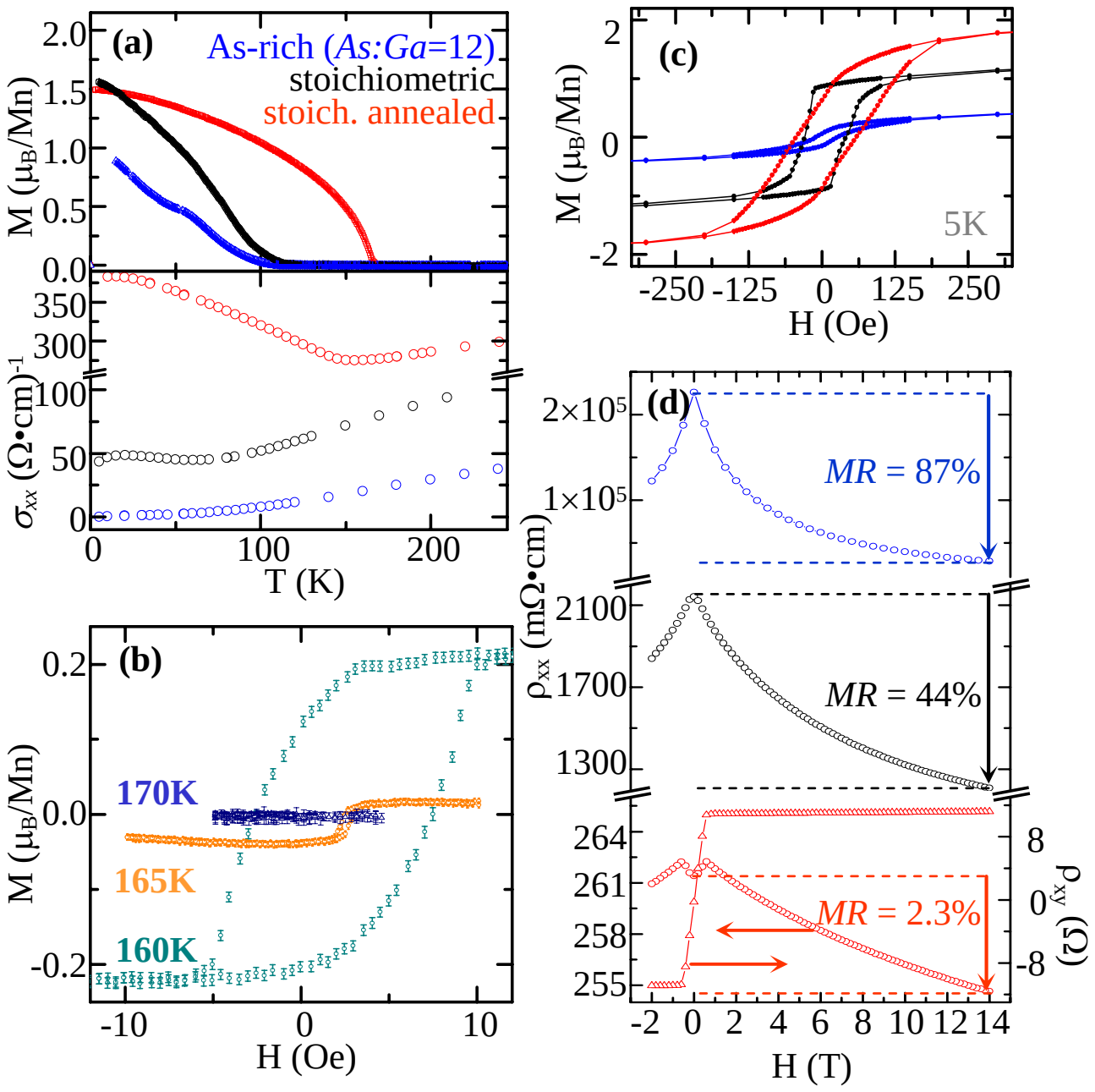

Figure 2
Mack et al.

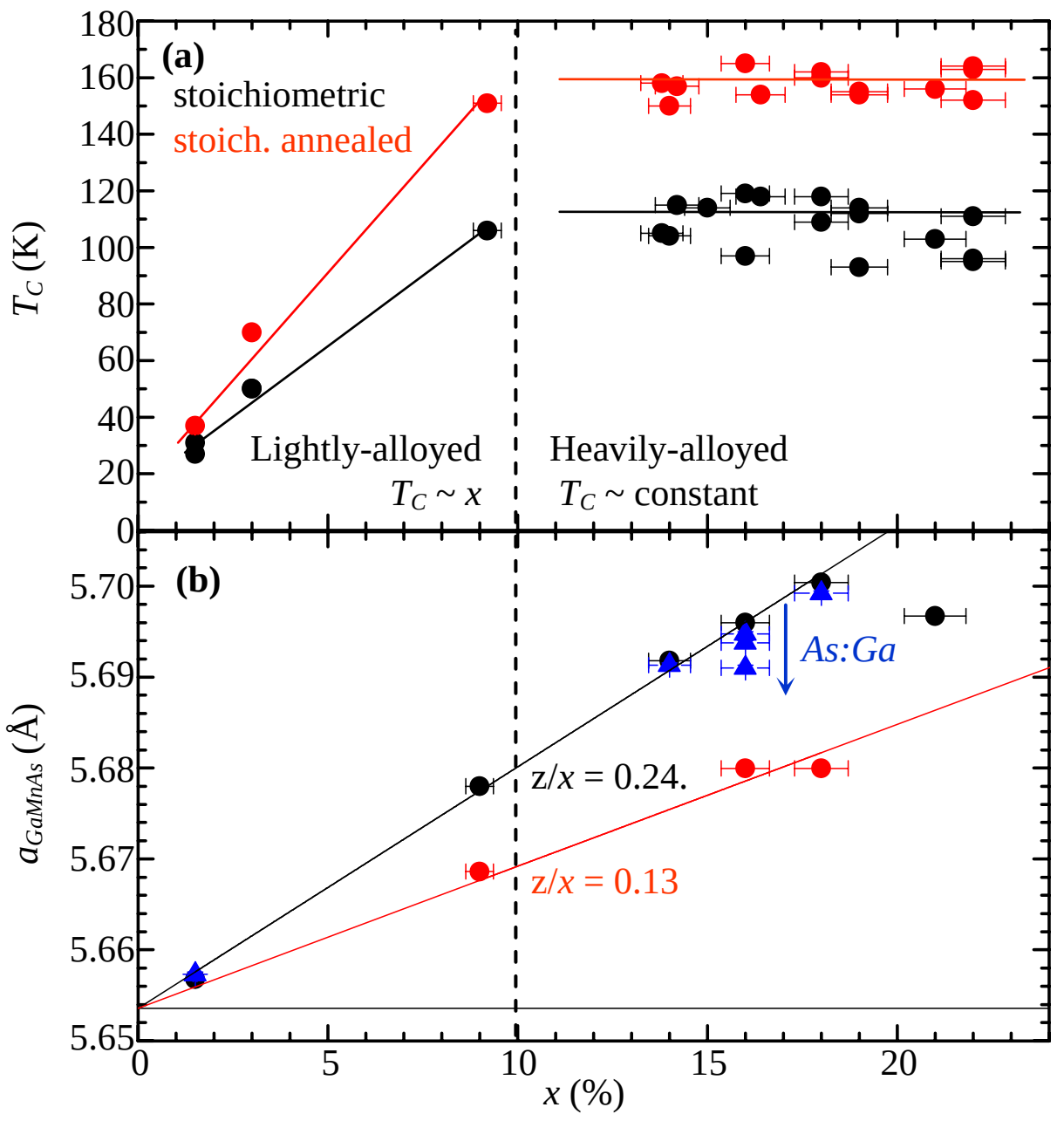

Figure 3
Mack et al.